\documentclass[pamm,a4paper,fleqn]{w-art}
\usepackage{times,cite,w-thm}
\usepackage[T1]{fontenc}
\usepackage[utf8]{inputenc}
\usepackage{hyperref}
\usepackage{graphicx}
\begin{document}

\TitleLanguage[EN]
\title[MaRDMO: FAIR Documentation of In-Silico Research]{MaRDMO: FAIR Documentation of In-Silico Research}

\author{\firstname{Marco} \lastname{Reidelbach}\inst{1,}%
\footnote{Corresponding author: e-mail \ElectronicMail{reidelbach@zib.de}}} 

\author{\firstname{Marcus} \lastname{Weber}\inst{1}%
    }

\address[\inst{1}]{\CountryCode[DE]Zuse Institute Berlin, Germany}

\AbstractLanguage[EN]
\begin{abstract}
MaRDMO is a plugin for the Research Data Management Organiser (RDMO) that enables the structured, FAIR-compliant documentation and discovery of mathematical research data. By embedding mathematics-specific questionnaires into a widely used data management plan tool, MaRDMO lowers the barrier to contributing and querying the MaRDI Knowledge Graph for researchers across all disciplines. This paper presents the current state of MaRDMO, including its questionnaire-driven documentation process for mathematical models, algorithms, and interdisciplinary workflows, illustrated through a concrete example based on a solver comparison study for the Stokes-Darcy system. We further describe the dedicated MaRDI RDMO instance as a ready-to-use entry point for the community, and discuss recent developments including a simplified Basic Model Catalog and an improved class-filtered search. The paper concludes with an outlook on LLM-assisted documentation features currently under development.
\end{abstract}
\maketitle                   

\section{Introduction}\label{sec:introduction}

Ask a mathematician whether they produce research data, and the answer is often a confident "no". This perspective is understandable. Mathematics has a long tradition of results expressed as theorems, proofs, and formulas rather than datasets or measurements\cite{Boege2023}. It does, however, create a practical challenge: research data management (RDM) support staff at universities and research institutions often find themselves uncertain how to assist mathematicians, unsure how to apply the tools and standards they routinely use with colleagues in other disciplines. Those standards, well-developed in many other disciplines\footnote{\url{https://rd-alliance.github.io/metadata-directory/}; \textit{Last accessed on May 13th, 2026.}}, have no direct equivalent in mathematics, and expectations around data management for mathematicians vary considerably across institutions and funding contexts\footnote{\url{https://www.math.harvard.edu/media/DataManagement.pdf}; \textit{Last accessed on April 22nd, 2026.}}.

\begin{figure}[ht]
    \centering
    \includegraphics[width=0.9\textwidth]{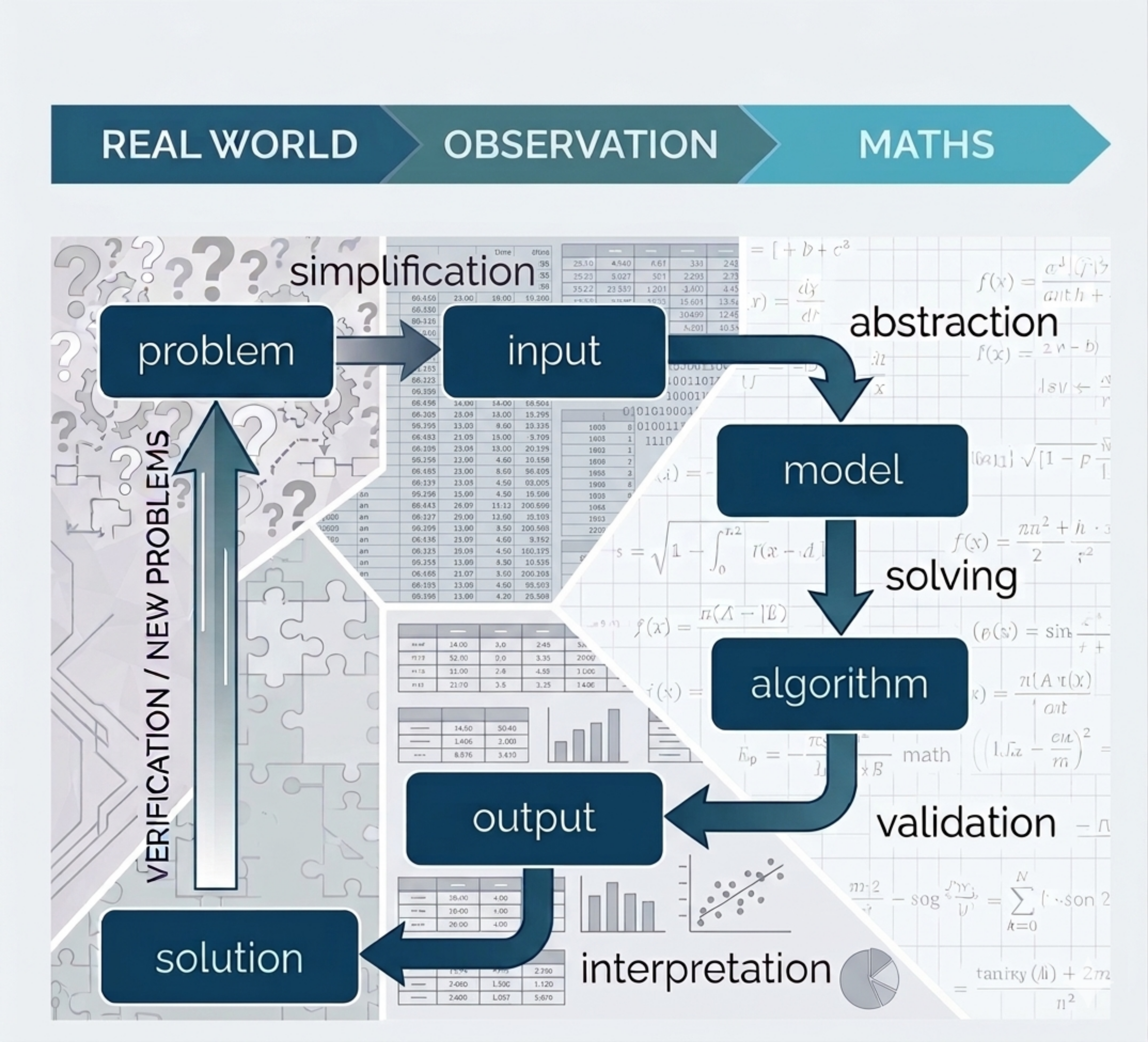}
    \caption{\small The model-simulation-optimization workflow underlying in-silico research. Real-world problems are simplified into inputs and abstracted into mathematical models, which are then solved by algorithms to produce outputs. These outputs are validated and interpreted to derive solutions, which in turn give rise to new problems and feed back into the cycle. Mathematical models, algorithms, and the data they generate and transform are the core research data types of this workflow. Diagram adapted from\cite{MaRDI2022} with Imagen 3 (accessed April 2026).}
    \label{fig:figure1}
\end{figure}

Yet mathematics is far from data-free. As illustrated in Figure \ref{fig:figure1}, the model-simulation-optimization cycle that underlies much of applied and computational mathematics is deeply data-generating: models abstract real-world problems, algorithms transform inputs into outputs, and results feed back into new hypotheses and refinements\cite{MaRDI2022}. This cycle, often referred to as in-silico research, pervades scientific disciplines far beyond mathematics itself. Mathematical models drive simulations in engineering and the natural sciences\cite{Alobaid2022}, algorithms underpin computational advances in medicine and economics\cite{Albuquerque2025}, and data-driven methods\cite{Alexander2023} across all disciplines rest on mathematical assumptions and structures. Making research data FAIR (Findable, Accessible, Interoperable, and Reusable\cite{Wilkinson2016}) in any of these fields therefore cannot be achieved without also making the underlying mathematical research data FAIR.

Recognizing this, the Mathematical Research Data Initiative, MaRDI, established within the German National Research Data Infrastructure, NFDI\cite{Hartl2021}, has set out to build the standards, services, and tools needed to bring the FAIR principles to mathematical research data. Central to this effort are ontologies and knowledge graphs for the three data types that appear most broadly across in-silico research: mathematical models, algorithms, and interdisciplinary workflows. The Mathematical Model Database, MathModDB\cite{Shehu2025, Schembera2025_CoRDI, Schembera2025, Schembera2024, Schembera2023}, provides a structured vocabulary and knowledge graph for documenting models, including their formulas, quantities, computational tasks, and the research problems and academic disciplines they address. Its counterpart, the Mathematical Algorithm Database, MathAlgoDB\cite{MathAlgoDB2022}, does the same for algorithms, capturing their relationships to algorithmic tasks, software implementations, and benchmarks. For interdisciplinary workflows, MaRDI has developed standardized documentation templates that record the full chain from research objective through processing steps, methods, software, hardware, and data\cite{Boege2023}. Together, these resources currently document 229 mathematical models, 512 algorithms, and a few interdisciplinary workflows, with ongoing community contributions expanding the collections\footnote{
\url{https://portal.mardi4nfdi.de/wiki/Project:KG_stats}; \textit{Statistics from May 13th, 2026.}}.

Having these resources is necessary but not sufficient. A knowledge graph that requires researchers to learn SPARQL, navigate Wikibase, and manually cross-reference multiple databases will see limited adoption regardless of its technical quality. Lowering these barriers is the purpose of the MaRDMO Plugin: a mathematics-specific extension of the Research Data Management Organiser, RDMO\cite{Neuroth2018}, that guides researchers through structured questionnaires to document and search mathematical research data without requiring any knowledge of the underlying technical infrastructure\cite{Reidelbach2026b, Reidelbach2026a, Reidelbach2025_CoRDI, Reidelbach_2025, Reidelbach2024, Reidelbach2023}. Because RDMO is already widely deployed across German research institutions and NFDI consortia, MaRDMO can integrate directly into existing local RDM processes rather than asking researchers or support staff to adopt an entirely new system\cite{Enke2023}.

This paper describes the current state of MaRDMO and the broader MaRDI ecosystem it connects to. Section~\ref{sec:the-mardi-ecosystem} introduces the MaRDI service landscape. Section~\ref{sec:the-mardmo-plugin} describes the MaRDMO Plugin, its data model, and its questionnaire-driven documentation process, illustrated through a concrete example. Section~\ref{sec:mardi-rdmo-instance} presents the new dedicated MaRDI RDMO instance as a ready-to-use entry point for the community. Sections~\ref{sec:recent-developments} and ~\ref{sec:outlook-and-conclusion} discuss recent developments, future directions, and conclusions.

\section{The MaRDI Ecosystem}\label{sec:the-mardi-ecosystem}

MaRDI has developed a broad portfolio of services addressing different aspects of mathematical RDM, spanning the full lifecycle from documentation and publication through execution and long-term reuse.

At the heart of the ecosystem is the MaRDI Portal, a Wikibase-powered web platform that hosts the MaRDI Knowledge Graph, a continuously growing, community-maintained knowledge base for mathematical research data\cite{Schubotz2023}. The MaRDI Knowledge Graph covers a wide range of mathematical entities including publications, software, proofs, and formulas, and serves as both a human-readable reference and a machine-queryable resource. Within this broader graph, MathModDB and MathAlgoDB form two dedicated subgraphs for the standardized description of mathematical models and algorithms, respectively. MathModDB documents models in terms of their (mathematical) formulas, involved quantities, associated computational tasks, and the research problems they address across various academic disciplines. MathAlgoDB provides an equivalent structured description of algorithms, linking them to the algorithmic tasks they solve, their implementing software, and relevant benchmarks. The connection between the computational tasks of MathModDB and the algorithmic tasks of MathAlgoDB enables cross-type queries that reflect the natural relationship between models and the algorithms used to solve them. Interdisciplinary workflows are documented as part of the overall MaRDI Knowledge Graph as well, based on standardized templates developed within MaRDI that draw in part on the Metadata4Ing ontology\cite{Iglezakis2023} developed by the NFDI4Ing consortium\cite{4ING}, capturing processing steps, methods, software, hardware, and datasets associated with specific research objectives.

Complementing the knowledge graph infrastructure, MaRDI provides a range of services targeting the practical needs of computational researchers. MaRDIFlow\cite{Pavan2025, Pavan2024, Pavan2023} offers a workflow framework for designing, executing, and documenting FAIR computational experiments, automatically capturing provenance and metadata and supporting multiple output formats for reporting and reproducibility. MaPS, the MaRDI Packaging System\cite{Kaushik2024}, addresses the challenge of software reproducibility by providing a unified interface for packaging software together with its complete dependency environment and deploying it on arbitrary systems. The MaRDI Open Interfaces project\cite{Kabanov2025} improves interoperability in scientific computing by providing automatic data marshalling between programming languages and standardized interfaces for common numerical problems such as differential equation integration and optimization, facilitating benchmarking and solver exchange. The mrdi file format\footnote{\url{https://portal.mardi4nfdi.de/wiki/Mrdi_File_Format_(Provide_format_for_storing_data_from_computer_algebra.)}; \textit{Last accessed on April 22nd, 2026.}} provides a JSON-based specification for storing and loading common data types in computer algebra, with implementations available in several major computer algebra systems.

A common thread across these services is a commitment to interoperability, both within the MaRDI ecosystem and within the broader research data landscape, ensuring that mathematical research data remains linkable rather than siloed. For researchers seeking a practical entry point into this ecosystem, whether to contribute documented models, algorithms, or interdisciplinary workflows, or to search and reuse existing ones, the MaRDMO Plugin provides the necessary bridge, and is described in detail in the following section.

\section{The MaRDMO Plugin}\label{sec:the-mardmo-plugin}

RDMO is an established web-based tool for creating data management plans (DMPs) and software management plans (SMPs), widely used across German research institutions and NFDI consortia. Its questionnaire and plugin architecture makes it highly adaptable to discipline-specific needs. The MaRDMO Plugin extends RDMO with mathematics-specific questionnaires and export functionality, connecting the familiar RDMO interface to the MaRDI Portal and Wikidata\cite{Vrandevic2014} as connected knowledge graphs. Through this extension, researchers can document mathematical models, algorithms, and interdisciplinary workflows in a standardized, ontology-driven format and publish them directly into the MaRDI Portal. In addition, a dedicated search catalog allows researchers to query the MaRDI Knowledge Graph for existing mathematical research data. The plugin and questionnaires are openly available, with the plugin installable as a Python package via PyPI\footnote{\url{https://pypi.org/project/MaRDMO/}; \textit{Last accessed on April 22nd, 2026.}} and the source code for both hosted on GitHub\footnote{\url{https://github.com/MaRDI4NFDI/MaRDMO-Plugin}; \textit{Last accessed on April 22nd, 2026.}}\footnote{\url{https://github.com/MaRDI4NFDI/MaRDMO-Questionnaire}; \textit{Last accessed on April 22nd, 2026.}}.

Each of the three supported data types has a dedicated questionnaire that reflects the structure of the corresponding ontology. All questionnaires share a common interaction pattern: for each ontology class, the researcher first searches for an existing item across the connected knowledge graphs. If a suitable item is found, the researcher selects it through MaRDMO, which automatically retrieves its metadata and pre-fills the relevant fields, reducing manual effort. The retrieved information can be extended by the researcher with additional metadata where needed. If no item exists in the connected knowledge graphs, the researcher provides the information manually, creating a new entry. This reuse-first approach minimizes redundant entries and actively builds on the existing knowledge base. A dedicated automation scheme ensures that selecting an item from one class triggers the automatic retrieval and population of all downstream related items, further reducing manual effort and enforcing completeness.

Publications can be linked to any item across all three documentation types. To support this, MaRDMO searches for publications not only in the connected knowledge graphs but also via the Crossref\footnote{\url{https://api.crossref.org/works/}}, DataCite\footnote{\url{https://api.datacite.org/dois/}}, zbMath\footnote{\url{https://api.zbmath.org/v1/document/_structured_search?page=0&results_per_page=100&DOI=}}, DOI\footnote{\url{https://citation.doi.org/metadata?doi=}}, and ORCID APIs\footnote{\url{https://pub.orcid.org/v3.0/search/?q=doi-self:}}, retrieving as much information as possible about the paper, its authors, and the publishing journal, in order to generate a proper publication entry in the MaRDI Portal.

Once the documentation is finalized, MaRDMO performs an (ontological) validation check before presenting a structured preview. The researcher can then export the documentation directly to the MaRDI Portal.

To illustrate these functionalities concretely, we use the Stokes-Darcy problem studied in Schmalfuss et al.\cite{Schmalfuss2021} as a running example across all three questionnaires. The problem describes the coupled flow of an incompressible fluid over a porous medium and is a representative model case for the challenges involved in solving coupled systems of partial differential equations.

\subsection{Documenting a Mathematical Model}\label{sec:document-model}

The model questionnaire is structured according to the MathModDB ontology and covers the model itself, its formulas, associated quantities and quantity kinds, computational tasks, research problems, academic disciplines, and scholarly publications.

For the Stokes-Darcy example, the researcher documents the continuous Stokes-Darcy model, which describes the coupled flow of an incompressible fluid over a porous medium. Let us assume that searching for this model in the connected knowledge graphs returns no existing entry, so the researcher defines it as a new model by providing a name and short description. The model properties are then specified: the model is deterministic, dimensional, dynamic, linear, space-continuous, and time-continuous.

For the associated research problem, the researcher searches the connected knowledge graphs and selects the existing entry "free flow coupled to porous media flow" found in the MaRDI Portal, whereupon MaRDMO automatically retrieves and pre-fills its name, description, and associated academic discipline. The researcher then specifies the model-to-model relations. Both the continuous Stokes model and the continuous Darcy model already exist in the MaRDI Portal, and the researcher selects them to declare that the new Stokes-Darcy model contains them. Since the two submodels are already fully documented, the only genuinely new content to provide are the formulas that couple them: the continuity of normal stresses, the continuity of normal mass fluxes, and the Beavers-Joseph-Saffman condition. These three formulas are not yet present in the connected knowledge graphs and are therefore defined as new entries. In doing so, the researcher can reuse several quantities already present in the MaRDI Portal, such as fluid velocity, fluid pressure, intrinsic permeability, dynamic viscosity, and the unit normal and tangent vectors, while two new quantities are introduced: the Beavers-Joseph coefficient and the viscous stress tensor. Table~\ref{tab:table1} summarizes the items involved in this documentation and their status.

\begin{table}[h]
\centering
\begin{tabular}{llll}
\hline
\textbf{Class} & \textbf{Item} & \textbf{Status} & \textbf{QID} \\
\hline
Mathematical Model & Stokes model & re-used & mardi:\href{https://portal.mardi4nfdi.de/wiki/Item:Q6675409}{Q6675409}\\
Mathematical Model & Darcy model & re-used & mardi:\href{https://portal.mardi4nfdi.de/wiki/Item:Q6675390}{Q6675390}\\
Mathematical Model & Stokes-Darcy model & newly created & -- \\
\hline
Formula & Stokes equations & re-used & mardi:\href{https://portal.mardi4nfdi.de/wiki/Item:Q6674516}{Q6674516} \\
Formula & Darcy equations & re-used & mardi:\href{https://portal.mardi4nfdi.de/wiki/Item:Q6674304}{Q6674304} \\
Formula & Continuity of normal stresses & newly created & -- \\
Formula & Continuity of normal mass fluxes & newly created & -- \\
Formula & Beavers-Joseph-Saffman condition & newly created & -- \\
\hline
Quantity & Fluid velocity (free flow) & re-used & mardi:\href{https://portal.mardi4nfdi.de/wiki/Item:Q6673788}{Q6673788} \\
Quantity & Fluid velocity (porous medium) & re-used & mardi:\href{https://portal.mardi4nfdi.de/wiki/Item:Q6673892}{Q6673892} \\
Quantity & Fluid pressure (free flow) & re-used & mardi:\href{https://portal.mardi4nfdi.de/wiki/Item:Q6673788}{Q6673788} \\
Quantity & Fluid pressure (porous medium) & re-used & mardi:\href{https://portal.mardi4nfdi.de/wiki/Item:Q6673894}{Q6673894} \\
Quantity & Fluid intrinsic permeability (porous medium) & re-used & mardi:\href{https://portal.mardi4nfdi.de/wiki/Item:Q6673787}{Q6673787} \\
Quantity & Kinematic viscosity & re-used & mardi:\href{https://portal.mardi4nfdi.de/wiki/Item:Q6673977}{Q6673977} \\
Quantity & Fluid mass density & re-used & mardi:\href{https://portal.mardi4nfdi.de/wiki/Item:Q6673976}{Q6673976} \\
Quantity & Fluid dynamic viscosity (porous medium) & re-used & mardi:\href{https://portal.mardi4nfdi.de/wiki/Item:Q6673913}{Q6673913} \\
Quantity & Dynamic viscosity & re-used & mardi:\href{https://portal.mardi4nfdi.de/wiki/Item:Q6673786}{Q6673786} \\
Quantity & Unit normal vector & re-used & mardi:\href{https://portal.mardi4nfdi.de/wiki/Item:Q6673790}{Q6673790} \\
Quantity & Unit tangent vector & re-used & mardi:\href{https://portal.mardi4nfdi.de/wiki/Item:Q6673791}{Q6673791} \\
Quantity & Beavers-Joseph coefficient & newly created & --\\
Quantity & Viscous stress tensor & newly created & --\\
\hline
Quantity Kind & Time & re-used & mardi:\href{https://portal.mardi4nfdi.de/wiki/Item:Q6534304}{Q6534304}\\
\hline
Research Problem & Free flow of an incompressible fluid & re-used & mardi:\href{https://portal.mardi4nfdi.de/wiki/Item:Q6684651}{Q6684651} \\
Research Problem & Flow in porous media & re-used & mardi:\href{https://portal.mardi4nfdi.de/wiki/Item:Q6684649}{Q6684649} \\
Research Problem & Free flow coupled to porous media flow & re-used & mardi:\href{https://portal.mardi4nfdi.de/wiki/Item:Q6684650}{Q6684650} \\
\hline
Academic Discipline & Continuum mechanics & re-used & mardi:\href{https://portal.mardi4nfdi.de/wiki/Item:Q6684700}{Q6684700} \\
\hline
\end{tabular}
\caption{Items involved in the documentation of the \textit{Stokes-Darcy model} using MaRDMO and their status in the MaRDI Portal. QIDs prefixed with \texttt{mardi:} refer to \url{https://portal.mardi4nfdi.de/wiki/Item:}, those prefixed with \texttt{wikidata:} to \url{https://www.wikidata.org/wiki/}.}
\label{tab:table1}
\end{table}

The interdisciplinary workflow studied here operates on a discretized form of this model, obtained by applying a first-order backward Euler scheme in time and finite volumes in space, yielding a time-discrete and space-discrete variant. This discretized model can be documented in the same way, with the continuous Stokes-Darcy model recorded as the model it discretizes, and the corresponding discrete versions of the Stokes and Darcy submodels selected from the MaRDI Portal as its components.

\subsection{Documenting an Algorithm}\label{sec:document-algorithm}

The algorithm questionnaire follows the MathAlgoDB ontology and covers the algorithm itself, the algorithmic tasks it addresses, implementing software, and benchmarks. The study of \textit{Schmalfuss et al.}~\cite{Schmalfuss2021} employs several solver and preconditioner combinations; we focus here on the \textit{Uzawa iteration}\cite{Uzawa1958}, which already exists as an entry in the MaRDI Portal with its algorithmic task (\textit{saddle point problems}, itself a specialization of \textit{linear systems of equations}). This makes it a suitable example for demonstrating how MaRDMO can be used not only to create new entries but also to expand existing ones.

Let us assume that the researcher searches for the \textit{Uzawa iteration} in the connected knowledge graphs, selects the existing entry found in the MaRDI Portal, and MaRDMO automatically retrieves and pre-fills its metadata. The researcher then extends the existing entry by adding \textit{DUNE-ISTL}\footnote{\url{https://github.com/dune-project/dune-istl}; \textit{Last accessed on April 22nd, 2026.}} as implementing software, the iterative solver template library in which the \textit{Uzawa iteration} is implemented in the context of this study. Through MaRDMO, an identifier of the information system for mathematical software (swmath:\href{https://zbmath.org/software/18749}{18749}\footnote{\texttt{swmath:} is a prefix for \url{https://zbmath.org/software/}.}) and the source code repository of \textit{DUNE-ISTL} can be directly added to the corresponding entry, making the software precisely identifiable and accessible to future users. In principle, a benchmark dataset characterizing the performance of the algorithm on specific task instances could be added in the same way, further enriching the entry for future users seeking to compare solver performance. Table~\ref{tab:table2} summarizes the items involved.

\begin{table}[h]
\centering
\begin{tabular}{llll}
\hline
\textbf{Class} & \textbf{Item} & \textbf{Status} & \textbf{QID}\\
\hline
Algorithm & Uzawa iteration & re-used & mardi:\href{https://portal.mardi4nfdi.de/wiki/Item:Q6825390}{Q6825390} \\
\hline
Algorithmic Task & Saddle point problems & re-used & mardi:\href{https://portal.mardi4nfdi.de/wiki/Item:Q6825764}{Q6825764} \\
Algorithmic Task & Linear systems of equations & re-used & mardi:\href{https://portal.mardi4nfdi.de/wiki/Item:Q6825573}{Q6825573} \\
\hline
Software & DUNE-ISTL & not found & -- \\
\hline
\end{tabular}
\caption{Items involved in the documentation of the \textit{Uzawa iteration} using MaRDMO and their status in the MaRDI Portal. QIDs prefixed with \texttt{mardi:} refer to \url{https://portal.mardi4nfdi.de/wiki/Item:}, those prefixed with \texttt{wikidata:} to \url{https://www.wikidata.org/wiki/}.}
\label{tab:table2}
\end{table}

\subsection{Documenting an Interdisciplinary Workflow}\label{sec:document-workflow}

The interdisciplinary workflow questionnaire captures the full research process through a sequence of processing steps, each associated with input and output data, algorithms, software, and hardware.

For the \textit{Solver Comparison Workflow for the Stokes-Darcy System}, let us assume that the researcher begins by searching for an existing workflow entry in the connected knowledge graphs. Finding no existing entry, the researcher defines it as a new workflow by providing a name and short description. The researcher then states the research objective and describes the overall workflow procedure. Here the research objective could be the \textit{analysis of the runtime and memory behavior of partitioned coupling and monolithic block-preconditioning approaches in comparison to direct solving}, motivated by the known limitations of sparse direct solvers: poor parallel scaling and trustworthiness issues under bad conditioning\cite{Schmalfuss2021}.

The researcher then selects the mathematical model underlying the workflow, where the discretized \textit{Stokes-Darcy model} in Section~\ref{sec:document-model} is found in the MaRDI Portal and selected. Subsequently, the researcher defines the processing steps of the workflow. The workflow comprises four processing steps:

\begin{itemize}\itemsep0pt
\item Partitioned coupling solving
\item Monolithic block-preconditioning solving
\item Direct solving
\item Comparison
\end{itemize}

Searching the connected knowledge graphs for the individual processing steps returns no existing entries, so all four are defined as new entries. Notably, the first three steps share a similar structure and can each serve as a template for the others within MaRDMO, allowing the researcher to reuse the documentation of one solving step as a starting point for the next, adjusting only the relevant algorithms and software. 

The researcher then searches, for each processing step, the connected knowledge graphs for existing dataset entries and selects them if found, or defines new ones if not. The first three processing steps all take the discretized linear system as input and produce a solved system together with runtime and memory measurements. The fourth and final step takes these outputs as input and produces the comparative runtime analysis that forms the central result of the study. For each dataset, a source or persistent identifier under which the data can be retrieved should be provided to ensure findability and reusability.

For each processing step, the researcher then selects the algorithms applied from the connected knowledge graphs, or creates new ones if not found. For algorithms not yet present in the MaRDI Portal, a minimal MathAlgoDB-conformant entry covering the algorithm, the algorithmic task it solves, and its implementing software is created directly within the workflow questionnaire. Where an algorithm is found in Wikidata, its metadata is automatically retrieved and used to pre-fill the new entry, reducing manual effort. The retrieved information can be extended by the researcher where needed. If a more thorough documentation is required, the dedicated algorithm catalog should be used instead, as demonstrated in Section~\ref{sec:document-algorithm}. For example, the \textit{Uzawa iteration} is found in the MaRDI Portal and selected directly, the \textit{Picard iteration}\cite{Picard1890} and the \textit{AMG method}\cite{Brandt1977} are found in Wikidata and their metadata is retrieved automatically, while \textit{PD-GMRES}\cite{Nunez2018} and \textit{Bi-CGSTAB}\cite{VanderVorst1992} are not found in either and are therefore created as new entries from scratch.

Likewise, for each algorithm the researcher selects or creates the software implementing it. \textit{preCICE}\footnote{\url{https://github.com/precice/precice}; \textit{Last accessed on April 22nd, 2026.}} (swmath:\href{https://zbmath.org/software/8713}{8713}) and \textit{UMFPACK}\footnote{\url{http://www.cise.ufl.edu/research/sparse/umfpack/}; \textit{Last accessed on April 22nd, 2026.}} (swmath:\href{https://zbmath.org/software/989}{989}) are not yet present in the MaRDI Portal but are found in Wikidata, so their metadata is retrieved automatically when they are selected. \textit{DuMux}\footnote{\url{https://github.com/dumux/dumux}; \textit{Last accessed on April 22nd, 2026.}} (swmath:\href{https://zbmath.org/software/14298}{14293}) is not found in any connected knowledge graph and is added entirely from scratch. \textit{DUNE-ISTL} is already present in the MaRDI Portal (c.f. Section~\ref{sec:document-algorithm}) and is linked directly. Both \textit{DuMux} and \textit{DUNE-ISTL} depend on \textit{DUNE}\footnote{\url{https://github.com/DUNE}; \textit{Last accessed on April 22nd, 2026.}} (swmath:\href{https://zbmath.org/software/1466}{1466}), which is found in Wikidata and recorded here as a software dependency for both. Table~\ref{tab:table3} summarizes all items involved in the workflow documentation and their status.

\begin{table}[h]
\centering
\begin{tabular}{llll}
\hline
\textbf{Class} & \textbf{Item} & \textbf{Status} & \textbf{QID} \\
\hline
Workflow & Comparing partitioned coupling vs.\ monolithic & newly created & -- \\
         & block-preconditioning for Stokes-Darcy systems & & \\
\hline
Processing Step & Partitioned coupling solving & newly created & -- \\
Processing Step & Monolithic block-preconditioning solving & newly created & -- \\
Processing Step & Direct solving & newly created & -- \\
Processing Step & Comparison & newly created & -- \\
\hline
Data Set & Discretized linear system & newly created & -- \\
Data Set & Solved system and runtime (partitioned) & newly created & -- \\
Data Set & Solved system and runtime (monolithic) & newly created & -- \\
Data Set & Solved system and runtime (direct) & newly created & -- \\
Data Set & Runtime comparison & newly created & -- \\
\hline
Model & Stokes-Darcy model (discretized) & re-used & created in Section~\ref{sec:document-model} \\
\hline
Algorithm & Uzawa iteration & re-used & mardi:\href{https://portal.mardi4nfdi.de/wiki/Item:Q6825390}{Q6825390} \\
Algorithm & Picard iteration & re-used & wikidata:\href{https://www.wikidata.org/wiki/Q1683631}{Q1683631} \\
Algorithm & Inverse least-squares interface quasi-Newton & re-used & wikidata:\href{https://www.wikidata.org/wiki/Q25098909}{Q25098909} \\
Algorithm & AMG method & re-used & wikidata:\href{https://www.wikidata.org/wiki/Q1471828}{Q1471828} \\
Algorithm & Block-Gauss-Seidel method & re-used & wikidata:\href{https://www.wikidata.org/wiki/Q1069090}{Q1069090} \\
Algorithm & ILU(0) factorization & re-used & wikidata:\href{https://www.wikidata.org/wiki/Q1654069}{Q1654069} \\
Algorithm & Unsymmetric MultiFrontal method & newly created & -- \\
Algorithm & PD-GMRES & newly created & -- \\
Algorithm & Bi-CGSTAB & newly created & -- \\
Algorithm & Block-Jacobi method & newly created & -- \\
\hline
Software & preCICE & re-used & wikidata:\href{https://www.wikidata.org/wiki/Q67123602}{Q67123602} \\
Software & DuMux & newly created & -- \\
Software & UMFPACK & re-used & wikidata:\href{https://www.wikidata.org/wiki/Q2467290}{Q2467290} \\
Software & DUNE-ISTL & re-used & created in Section~\ref{sec:document-algorithm} \\
Software & DUNE & re-used & wikidata:\href{https://www.wikidata.org/wiki/Q1265582}{Q1265582} \\
\hline
Hardware & AMD EPYC 7551P CPU & newly created & -- \\
\hline
\end{tabular}
\caption{Items involved in the documentation of the \textit{Solver Comparison Workflow for the Stokes-Darcy System} using MaRDMO and their status in the MaRDI Portal. QIDs prefixed with \texttt{mardi:} refer to \url{https://portal.mardi4nfdi.de/wiki/Item:}, those prefixed with \texttt{wikidata:} to \url{https://www.wikidata.org/wiki/}.}
\label{tab:table3}
\end{table}

\subsection{Searching the MaRDI Knowledge Graph}\label{sec:search-the-graph}

Beyond documentation, MaRDMO provides a dedicated search catalog that allows researchers to query the MaRDI Knowledge Graph without any knowledge of SPARQL. Rather than contributing new entries, the search catalog goes in the opposite direction: the researcher specifies search criteria through a guided questionnaire interface, and MaRDMO translates these into a SPARQL query that is executed against the MaRDI Portal on the researcher's behalf. The resulting query and its results are presented directly in the RDMO interface.
The search catalog supports queries across all three documented data types. Researchers can search for mathematical models based on the research problems they address or the computational tasks they are applied to, find algorithms based on the tasks they solve or the software in which they are implemented, and discover interdisciplinary workflows based on specific research objectives, applied methods, or software tools. This allows researchers from non-mathematical disciplines to explore relevant mathematical research data based on functional criteria, without requiring detailed knowledge of the underlying mathematical theory or the structure of the knowledge graph.

\subsection{Export and Search}\label{sec:export-and-search}

Once a documentation catalog is completed, the researcher can initiate the export via the "Export to MaRDI Portal" button, which MaRDMO adds to the RDMO export section. Before the actual export, MaRDMO generates a structured preview summarizing all documented information. During this step, MaRDMO performs a series of checks to ensure the documentation is complete, consistent, and Wikibase-conformant. Completeness is verified by checking that all items have their required downstream items: for example, a documented research problem must be linked to an academic discipline, a computational task must be connected to formulas and quantities, and formulas must in turn be linked to quantities. Consistency checks identify logical contradictions, such as a formula being marked as both linear and nonlinear simultaneously. Wikibase-related checks flag issues such as short descriptions exceeding the maximum allowed length. Problematic aspects are indicated to the researcher, who can return to the questionnaire to address them before proceeding.

If the documentation passes all checks, the researcher can confirm the export by clicking "Export to MaRDI Portal" at the end of the preview page. All three data types are exported to the MaRDI Portal via the REST API following the OAuth2 authentication protocol using the researcher's institutional credentials, ensuring that newly created items are attributed to the contributing researcher. While MaRDMO adds machine-readable information to the knowledge graphs, human-readable representations are automatically generated through structured templates on each platform. Once the export is completed, MaRDMO presents a landing page with direct links to the newly created items. To maintain data quality beyond automated checks, a second validation step must be carried out by domain experts from the individual service providers, who review contributed entries for accuracy and relevance.

For the search catalog, the researcher initiates the process via the "Query MaRDI Portal" button. MaRDMO first generates the corresponding SPARQL query from the researcher's input and displays it, giving researchers the opportunity to inspect and learn from the query. Upon confirming by clicking "Query MaRDI Portal" again, the SPARQL query is executed against the MaRDI Knowledge Graph and the results are presented to the researcher with direct links to the relevant entries in the MaRDI Portal.

\section{The MaRDI RDMO Instance}\label{sec:mardi-rdmo-instance}

The MaRDMO Plugin and questionnaires can be integrated into any existing RDMO instance, allowing researchers at universities and research institutions that already operate an RDMO instance to use MaRDMO without leaving their familiar local RDM environment. For all others, MaRDI provides a dedicated RDMO instance \footnote{\url{https://rdmo.mardi4nfdi.de/}; \textit{Last accessed on April 22nd, 2026.}}, removing the need to install software or manage any infrastructure. Researchers can start documenting immediately after logging in.

The instance is provided through the NFDI basic service DMP4NFDI\cite{Diederichs2024} and is hosted at ULB Darmstadt as part of a multisite platform serving several NFDI consortia. Login is available through NFDI AAI\footnote{\url{https://www.nfdi-aai.de/}; \textit{Last accessed on April 22nd, 2026.}} with Didmos or Unity, making the instance accessible to researchers from any discipline using their institutional login or ORCID. As the platform serves multiple consortia, MaRDMO can in principle be rolled out directly to the RDMO instances of other NFDI consortia in the future, bringing mathematical RDM services to researchers in their own disciplinary environment rather than requiring them to use a separate instance. This reflects the vision of a unified NFDI in which mathematical services are available across all disciplines as part of \textit{One NFDI}.

The instance comes, so far, with the documentation catalog for mathematical models described in Section~\ref{sec:the-mardmo-plugin}. In addition, a \textit{Basic Model Catalog} is available, offering a simplified entry point for model documentation that is introduced in Section~\ref{sec:recent-developments}. The algorithm, interdisciplinary workflow, and search catalogs described in Section~\ref{sec:the-mardmo-plugin} are available on GitHub\footnote{\url{https://github.com/MaRDI4NFDI/MaRDMO-Questionnaire}; \textit{Last accessed on April 22nd, 2026.}} and will be made available on the instance upon an upcoming update. Beyond the MaRDMO-specific catalogs, the instance also provides an SMP template\footnote{\url{https://github.com/rdmorganiser/rdmo-catalog/blob/main/rdmorganiser/questions/questions-smp.xml}; \textit{Last accessed on April 22nd, 2026.}} developed by the \textit{Max Planck Digital Library} and a generic DMP template\footnote{\url{https://github.com/rdmorganiser/rdmo-catalog/blob/main/rdmorganiser/questions/questions-DFG-Checkliste.xml}; \textit{Last accessed on April 22nd, 2026.}}, allowing researchers to address all their data and software management planning needs in one place. A mathematics-specific DMP template is currently under development in an incubator project\footnote{\url{https://dmp.services.base4nfdi.de/incubator}; \textit{Last accessed on April 22nd, 2026.}} together with DMP4NFDI.

The MaRDI RDMO instance is accompanied by several support and community resources. Questions, feedback, and feature requests can be directed to the MaRDI RDMO team\footnote{\href{mailto:rdmo@mardi4nfdi.de}{rdmo@mardi4nfdi.de}}, and within each questionnaire a dedicated button allows researchers to send a message directly in the context of a specific question. A mailing list\footnote{\href{mailto:mardi-rdmo@listserv.dfn.de}{mardi-rdmo@listserv.dfn.de}} provides a forum for questions and discussions around the instance, RDMO in general, MaRDMO, as well as DMPs and SMPs, and is also used to share news and updates. A monthly online meeting open to all interested researchers provides a regular opportunity for discussion. Finally, YouTube videos demonstrating the documentation of mathematical models\footnote{\url{https://www.youtube.com/playlist?list=PLgoPZ7uPWbo-jqDXzx4fSm_4JyAYEMPjn}; \textit{Last accessed on April 22nd, 2026.}} and algorithms\footnote{\url{https://www.youtube.com/playlist?list=PLgoPZ7uPWbo-aC9pnVMYRZYM3iygBYWwn}; \textit{Last accessed on April 22nd, 2026.}} through MaRDMO are available to help researchers get started.

Together, these resources lower the barrier to entry for researchers across all disciplines who wish to engage with FAIR mathematical RDM, whether they are documenting their own work for the first time or looking to discover and reuse existing mathematical models, algorithms, and interdisciplinary workflows from the MaRDI ecosystem.

\section{Recent Developments}\label{sec:recent-developments}

\begin{figure}[ht]
    \centering
    \includegraphics[width=0.9\textwidth]{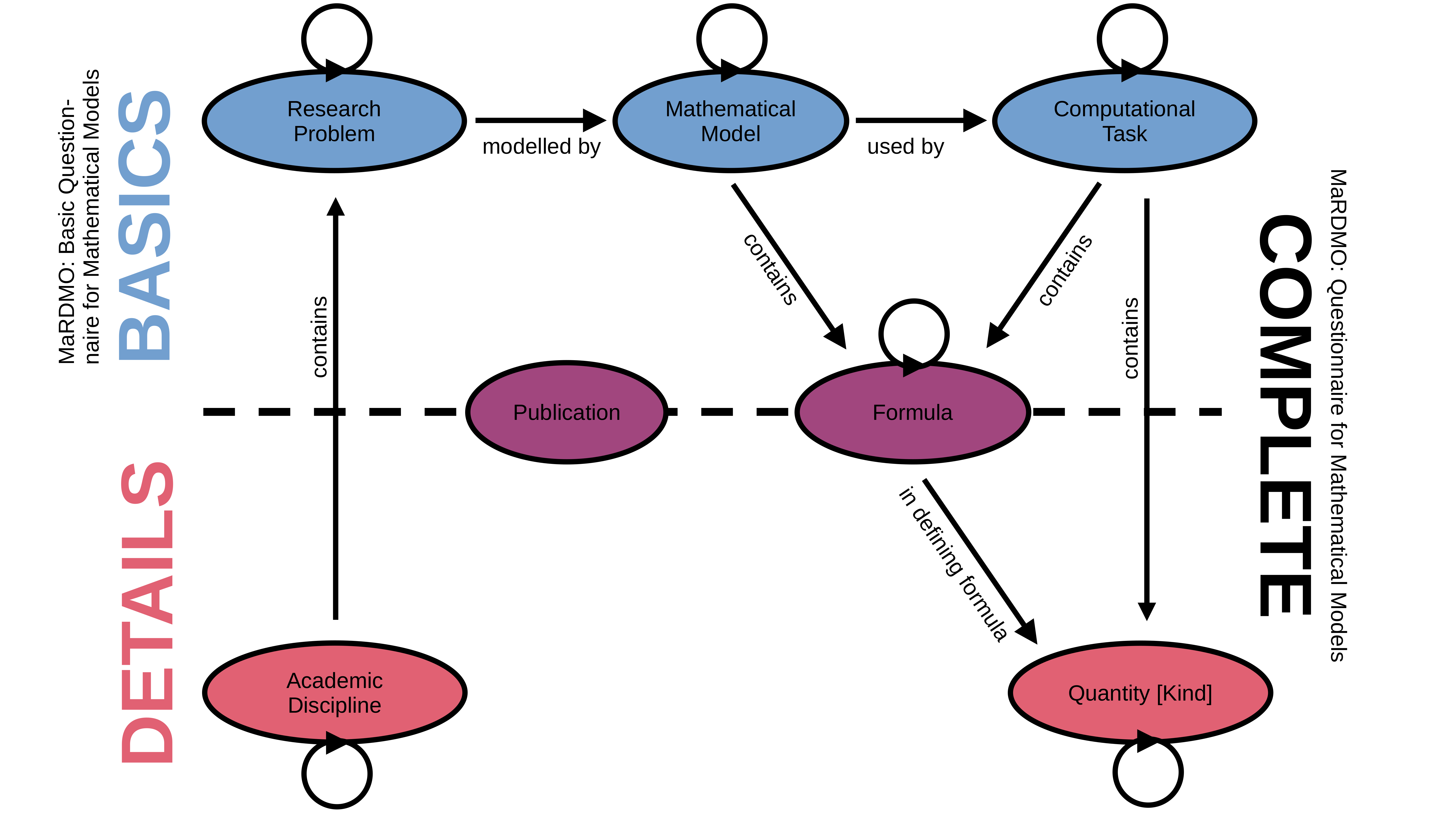}
    \caption{\small Structure of the \textit{Basic Model Catalog} (top) and the \textit{Complete Model Catalog} (bottom) for mathematical models in MaRDMO. The basic catalog covers the core classes Research Problem, Mathematical Model, and Computational Task, with formulas represented as lightweight placeholder entries rather than full MathModDB-conformant descriptions including quantities and quantity kinds.}
    \label{fig:figure2}
\end{figure}

Two recent developments extend the functionality of MaRDMO. The first is the \textit{Basic Model Catalog}, which reduces the required input for model documentation to the essential structural information, lowering the barrier for researchers documenting a model for the first time. As illustrated in Figure~\ref{fig:figure2}, the basic catalog covers three core classes: the mathematical model itself, including its properties and optional relations to other models, the research problem it models, and the computational tasks by which it is used. Formulas can still be linked to models and tasks, but rather than requiring the full \LaTeX representation with all associated quantities and quantity kinds, the researcher only needs to provide a meaningful reference to the formula, such as a DOI and equation number, creating a placeholder entry in the knowledge graph that can be completed at a later stage. Academic disciplines, quantities, and quantity kinds are not part of the basic catalog, further simplifying the process. The overall reuse-first approach is maintained: the connected knowledge graphs are searched before any new item is created, and existing entries are reused and extended wherever possible. Once a researcher is ready to provide a more thorough documentation, the \textit{Complete Model Catalog} can be used to enrich the existing entry with the missing details.

The second development concerns the item search within MaRDMO. Previously, item search relied on the unfiltered MediaWiki search API, which returned autocomplete results from all item classes simultaneously, making it difficult for researchers to identify relevant entries efficiently. A recent improvement addresses this by leveraging the MediaWiki search API with a Wikibase-specific \texttt{haswbstatement} filter, which restricts search results to items carrying a specific instance-of statement, effectively scoping the search to a particular ontology class. This means that when a researcher searches for a mathematical model, only items classified as mathematical models are returned, and similarly for research problems, algorithms, software, and other classes. The improvement in precision and usability is significant, as researchers now get a much more targeted view of what is already present in the knowledge graph, directly supporting the reuse-first approach that is central to MaRDMO. 

\section{Outlook and Conclusion}\label{sec:outlook-and-conclusion}

Looking ahead, two LLM-assisted developments are being pursued to further reduce the manual effort involved in using MaRDMO. The first is the integration of the RDMO chatbot, recently released by the RDMO development team, into the MaRDI RDMO instance. The vision is to configure this chatbot in a context-aware way, such that it acts as a DMP or SMP expert when the researcher is working on a DMP or SMP, a MathModDB expert when documenting a mathematical model, a MathAlgoDB expert when documenting an algorithm, and so on, providing researchers with immediate and contextually relevant guidance throughout the documentation process. The second development is an import plugin that allows researchers to provide a scholarly publication as input and uses a large language model to extract the relevant model, algorithm, or interdisciplinary workflow information and map it onto the corresponding MaRDMO questionnaire. The result is a version-zero documentation that the researcher can then review, adjust, extend, or partially discard as appropriate. This human-in-the-loop approach has the potential to dramatically reduce the effort required for documentation, in particular for researchers who are new to structured RDM, while ensuring that the final documented entry reflects the researcher's own judgment and domain expertise. Beyond AI-assisted features, MaRDMO will also be extended with additional catalogs to cover further MaRDI services. A dedicated catalog to accompany \textit{MaRDIFlow} is planned, supporting the documentation of FAIR computational experiments within the familiar MaRDMO interface. Furthermore, links to \textit{MaPS} and \textit{MaRDI Open Interfaces} will be established within the interdisciplinary workflow catalog, allowing researchers to reference reproducible software environments and standardized solver interfaces directly in their workflow documentation.

MaRDMO has established itself as a practical gateway between researchers and the MaRDI ecosystem, making the documentation and discovery of mathematical research data accessible without requiring expertise in knowledge graphs, ontologies, or SPARQL. By embedding this functionality within RDMO, a tool already familiar to many researchers and RDM support staff, MaRDMO lowers the barrier to FAIR documentation of in-silico research to a level that is realistic for everyday research practice. The \textit{Stokes-Darcy} solver comparison example presented in this paper illustrates how a single research study naturally spans all three documentation types, and how the reuse-first approach of MaRDMO allows researchers to build on an already growing body of documented mathematical research data rather than starting from scratch. As the MaRDI Knowledge Graph continues to grow through community contributions, the value of each new documented entry increases: models, algorithms, and interdisciplinary workflows become more findable, more reusable, and more connected across disciplines. Making in-silico research FAIR is not a one-time effort but a cumulative one, and MaRDMO is designed to make each individual contribution as effortless as possible.

\vspace{\baselineskip}
 \bibliographystyle{pamm}
  \bibliography{PAMM}

\end{document}